\documentstyle[12pt]{article}

\makeatletter

\newif\ifnoncomplete

\def\final{\noncompletefalse\typeout{** FINAL form (substyle:Sachiko)}}
\noncompletefalse

\newif\ifrefphysrev

\refphysrevfalse

\def \vol(#1,#2,#3){\ifrefphysrev{{\bf {#1}},
{#3} (19{#2})}\else{{{\bf {#1}}(19{#2}){#3}}}\fi}
\def \NP(#1,#2,#3){Nucl.\ Phys.\          \vol(#1,#2,#3)}
\def \PL(#1,#2,#3){Phys.\ Lett.\          \vol(#1,#2,#3)}
\def \PRL(#1,#2,#3){Phys.\ Rev.\ Lett.\   \vol(#1,#2,#3)}
\def \PRp(#1,#2,#3){Phys.\ Rep.\          \vol(#1,#2,#3)}
\def \PR(#1,#2,#3){Phys.\ Rev.\           \vol(#1,#2,#3)}
\def \PTP(#1,#2,#3){Prog.\ Theor.\ Phys.\ \vol(#1,#2,#3)}
\def \ibid(#1,#2,#3){{\it ibid.}\         \vol(#1,#2,#3)}

\def\thebibliography#1{
\section*{References\@mkboth
  {REFERENCES}{REFERENCES}}\list
  {[\arabic{enumi}]}{\setlength\labelwidth{2ex}
   \setlength\labelsep{0.05in} 
   \setlength\leftmargin{0.25in}
    \setlength\itemsep{0pt}
    \setlength\parsep{0pt}
   \itemsep\parskip
    \usecounter{enumi}}
    \def\newblock{\hskip .11em plus .33em minus -.22em}
    \sloppy
    \sfcode`\.=1000\relax}

\def\@bibitem#1{\item\if@filesw \immediate\write\@auxout
       {\string\bibcite{#1}{\the\c@enumi}}\fi\ignorespaces
       {\ifnoncomplete\reversemarginpar{\hspace*{-1.05in}\makebox[1in][l]
       {{\footnotesize{\sl [#1]}}}}\fi}%
       }

\def\@cite#1#2{\unskip\nobreak\relax
    {[#1]}} 

\def\citenum#1{{\def\@cite##1##2{##1}\cite{#1}}}
\def\citea#1{\@cite{#1}{}}

\newcount\@tempcntc
\def\@citex[#1]#2{\if@filesw\immediate\write\@auxout{\string\citation{#2}}\fi
  \@tempcnta\z@\@tempcntb\m@ne\def\@citea{}\@cite{\@for\@citeb:=#2\do
    {\@ifundefined
       {b@\@citeb}{\@citeo\@tempcntb\m@ne\@citea\def\@citea{,}{\bf ?}\@warning
       {Citation `\@citeb' on page \thepage \space undefined}}%
    {\setbox\z@\hbox{\global\@tempcntc0\csname b@\@citeb\endcsname\relax}%
     \ifnum\@tempcntc=\z@ \@citeo\@tempcntb\m@ne
       \@citea\def\@citea{,}\hbox{\csname b@\@citeb\endcsname}%
     \else
      \advance\@tempcntb\@ne
      \ifnum\@tempcntb=\@tempcntc
      \else\advance\@tempcntb\m@ne\@citeo
      \@tempcnta\@tempcntc\@tempcntb\@tempcntc\fi\fi}}\@citeo}{#1}}
\def\@citeo{\ifnum\@tempcnta>\@tempcntb\else\@citea\def\@citea{,}%
  \ifnum\@tempcnta=\@tempcntb\the\@tempcnta\else
   {\advance\@tempcnta\@ne\ifnum\@tempcnta=\@tempcntb \else \def\@citea{--}\fi
    \advance\@tempcnta\m@ne\the\@tempcnta\@citea\the\@tempcntb}\fi\fi}

\def\affiliation#1{\cr
\makebox[0in]{\parbox{8in}{\begin{center} {\sl #1}\end{center}}} \cr}
\def\@affiliation{}

\def\and{\cr \makebox[0in]{\rule[-1cm]{0mm}{1cm}and } \cr}

\def\maketitle{\par
 \begingroup
 \def\thefootnote{\fnsymbol{footnote}}
 \def\@makefnmark{\hbox
 to 0pt{$^{\@thefnmark}$\hss}}
 \if@twocolumn
 \twocolumn[\@maketitle]
 \else \newpage
 \global\@topnum\z@ \@maketitle \fi\thispagestyle{plain}\@thanks
 \endgroup
 \setcounter{footnote}{0}
 \let\maketitle\relax
 \let\@maketitle\relax
 \gdef\@thanks{}\gdef\@author{}\gdef\@title{}
 \gdef\@affiliation{} \let\affiliation\relax	%
 \let\thanks\relax}

\def\@maketitle{\newpage
 \null
 \vskip 0em plus 2em minus 0em     
 \ifx\@date\@empty\else
   \begin{flushright}
    {\ifnoncomplete(\today)
     \else{{\normalsize \@date}\\}\fi}      
   \end{flushright}
   \vskip 3em plus 2em minus 2em   
 \fi
 \begin{center}
  {\frtnsfb \@title \par}     
  \vskip 3em plus 1em minus 1.5em  
  {
   \lineskip .5em plus 0em minus .3em   
   \begin{tabular}[t]{c}\@author
   \end{tabular}\par}
\end{center}
 \par
 \vskip 6em plus 2em minus 4em}     

\def\abstract{\if@twocolumn
\section*{Abstract}
\else \normalsize
\fi}

\def\endabstract{\if@twocolumn\fi\par\clearpage}

\def\section{\@startsection {section}{1}{\z@}{3.5ex plus 1ex minus
    .2ex}{2.3ex plus .2ex}{\normalsize\bf}}
\def\subsection#1{\subsectioncom{\sc{#1}}}
\def\subsectioncom{\@startsection{subsection}{2}{\z@}
    {3.25ex plus 1ex minus .2ex}{1.5ex plus .2ex}{\small}}
\def\subsubsection{\@startsection{subsubsection}{3}{\z@}{3.25ex plus
1ex minus .2ex}{1.5ex plus .2ex}{\small}}

\def\@addmarginpar{\@next\@marbox\@currlist{\@cons\@freelist\@marbox
    \@cons\@freelist\@currbox}\@latexbug\@tempcnta\@ne
    \if@twocolumn
        \if@firstcolumn \@tempcnta\m@ne \fi
    \else
      \if@mparswitch
         \ifodd\c@page \else\@tempcnta\m@ne \fi
      \fi
      \if@reversemargin \@tempcnta -\@tempcnta \fi
    \fi
    \ifnum\@tempcnta <\z@  \global\setbox\@marbox\box\@currbox \fi
    \@tempdima\@mparbottom \advance\@tempdima -\@pageht
       \advance\@tempdima\ht\@marbox \ifdim\@tempdima >\z@
      \else\@tempdima\z@ \fi
    \global\@mparbottom\@pageht \global\advance\@mparbottom\@tempdima
       \global\advance\@mparbottom\dp\@marbox
       \global\advance\@mparbottom\marginparpush
    \advance\@tempdima -\ht\@marbox
    \global\ht\@marbox\z@ \global\dp\@marbox\z@
    \vskip -\@pagedp \vskip\@tempdima\nointerlineskip
    \hbox to\columnwidth
      {\ifnum \@tempcnta >\z@
          \hskip\columnwidth \hskip\marginparsep
        \else \hskip -\marginparsep \hskip -\marginparwidth \fi
       \box\@marbox \hss}
    \vskip -\@tempdima
    \nointerlineskip
    \hbox{\vrule \@height\z@ \@width\z@ \@depth\@pagedp}}

\def\ref#1{
    \@ifundefined{r@#1}{{#1}\@warning{Reference `#1'
    on page \thepage \space
    undefined}}{\edef\@tempa{\@nameuse{r@#1}}\expandafter
    \@car\@tempa \@nil\null}}
\def\refn#1{\@ifundefined{r@#1}{{#1}\@warning{Reference `#1'
    on page \thepage \space
    undefined}}{\edef\@tempa{\@nameuse{r@#1}}\expandafter
    \@car\@tempa \@nil\null}}

\def\endequationl{\eqno \@eqnnum 
$$\global\@ignoretrue}

\def\eqnarray{\stepcounter{equation}\let\@currentlabel=\theequation
\global\@eqnswtrue
\global\@eqcnt\z@\tabskip\@centering\let\\=\@eqncr
$$\arraycolsep\z@
\halign to \displaywidth\bgroup\@eqnsel\hskip\@centering
  $\displaystyle\tabskip\z@{##}$&\global\@eqcnt\@ne
  \hskip 2\arraycolsep \hfil$\displaystyle{{}##{}}$\hfil
  &\global\@eqcnt\tw@ \hskip 2\arraycolsep
  $\displaystyle\tabskip\z@{##}$\hfil
   \tabskip\@centering&\llap{##}\tabskip\z@\cr}
\def\mmodetrue{\mmode=\iftrue}

\def\eqnarrayl#1{\stepcounter{equation}\let\@currentlabel=\theequation
\label {#1}
\global\@eqnswtrue
\global\@eqcnt\z@\tabskip\@centering\let\\=\@eqncr
$$\arraycolsep\z@
\halign to \displaywidth\bgroup\@eqnsel\hskip\@centering
  $\displaystyle\tabskip\z@{##}$&\global\@eqcnt\@ne
  \hskip 2\arraycolsep \hfil$\displaystyle{{}##{}}$\hfil
  &\global\@eqcnt\tw@ \hskip 2\arraycolsep
  $\displaystyle\tabskip\z@{##}$\hfil
   \tabskip\@centering&\llap{##}\tabskip\z@\cr}
\def\label#1{
\@bsphack\if@filesw {
{\ifnoncomplete{\makebox[1in][r]{\footnotesize{\sl [#1]}}}\fi}%
\let\thepage\relax
   \xdef\@gtempa{\write\@auxout{\string
      \newlabel{#1}{{\@currentlabel}{\thepage}}}}
}\@gtempa
   \if@nobreak \ifvmode\nobreak\fi\fi\fi\@esphack}

\def\newlabel#1#2{
\@ifundefined{r@#1}{}{\@warning{Label `#1' multiply
   defined}}\global\@namedef{r@#1}{#2}}

\def\endeqnarrayl{\@@eqncr\egroup
      \global\advance\c@equation\m@ne$$\global\@ignoretrue}

\newif\if@numbersec \@numbersectrue
\def\appendix{\par\clearpage
  \setcounter{section}{0}
  \setcounter{subsection}{0}
  \def\thesection{\Alph{section}}
  \def\thesubsection{\arabic{subsection}}
  \@ifstar{\def\@sectname{Appendix}\@numbersecfalse}
          {\def\@sectname{Appendix~}\@numbersectrue}}

\def\thefigures#1{\par\clearpage\section*{Figures\@mkboth
  {FIGURES}{FIGURES}}\list
  {Fig.~\arabic{enumi}.}{\labelwidth\parindent\advance\labelwidth -\labelsep
      \leftmargin\parindent\usecounter{enumi}}}

\def\thetables#1{\par\clearpage\section*{Tables\@mkboth
  {TABLES}{TABLES}}\list
  {Table~\arabic{enumi}.}{\labelwidth-\labelsep
      \leftmargin0pt\usecounter{enumi}}}

\def\@sect#1#2#3#4#5#6[#7]#8{\ifnum #2>\c@secnumdepth
     \def\@svsec{}\else
     \refstepcounter{#1}\edef\@svsec{\ifnum #2=1 \@sectname
         \if@numbersec\csname the#1\endcsname\fi.\else
         \csname the#1\endcsname.\fi
        \hskip 1em }\fi
     \@tempskipa #5\relax
      \ifdim \@tempskipa>\z@
        \begingroup #6\relax
          \@hangfrom{\hskip #3\relax\@svsec}{\interlinepenalty \@M #8\par}
        \endgroup
       \csname #1mark\endcsname{#7}\addcontentsline
         {toc}{#1}{\ifnum #2>\c@secnumdepth \else
                      \protect\numberline{\csname the#1\endcsname}\fi
                    #7}\else
        \def\@svsechd{#6\hskip #3\@svsec #8\csname #1mark\endcsname
                      {#7}\addcontentsline
                           {toc}{#1}{\ifnum #2>\c@secnumdepth \else
                             \protect\numberline{\csname the#1\endcsname}\fi
                       #7}}\fi
     \@xsect{#5}}

\def\@sectname{}

 \def\thefootnote{\fnsymbol{footnote}}

\def \@magscale#1{ scaled \magstep #1}
\font\frtnsfb = cmssbx10 \@magscale2 
\newcount\ieq
\newcount\jeq
\newcount\keq
\def \eq{
\multiply\ieq by 2
\jeq=\ieq
\divide\jeq by 4
\multiply\jeq by 4
\ifnum\ieq=\jeq \end{eqnarray} \keq=1 
\else
\keq=2 \begin{eqnarray} \fi
\ieq=\keq
}

\def \mathbox(#1){\invisible\ifmmode{{#1}}\else{\mbox{${#1}$}}\fi}
\def \mbf(#1){\mbox{\boldmath{$#1$}}}
\def \abs(#1){\mathbox(\left|{#1}\right|)}
\def \bracket(#1){\mathbox(\left\langle{#1}\right\rangle)}
\def \brav(#1){\mathbox(\langle {#1}|)}
\def \cg(#1,#2,#3,#4,#5,#6){\mathbox({(#1\,#2\,#3\,#4|#5\,#6)})}
\def \comm(#1,#2){\mathbox(\left[{#1},{#2}\right])}
\def \dfdx(#1,#2){\mathbox(\frac{{\rm d}{#1}}{{\rm d}{#2}})}
\def \delfdelx(#1,#2){\mathbox(\frac{\partial{#1}}{\partial{#2}})}
\def \inprod(#1,#2){\mathbox({(#1\cdot #2)})}
\def \inprodij(#1){\mathbox({\inprod(#1_i,#1_j)})}
\def \intd(#1,#2){\mathbox({\int^#1_#2 \; \rmd})}
\def \eps(#1){\mathbox(\epsilon_{#1})}
\def \half(#1){\mathbox(\frac{#1}{2})}

\def \ketv(#1){\mathbox(|{#1}\rangle)}
\def \matele(#1,#2,#3){\mathbox(\left\langle {#1}|\,{#2}\,|{#3}\right\rangle)}
\def \mateled(#1,#2,#3){\mathbox(\left\langle
{#1}||\,{#2}\,||{#3}\right\rangle)}
\def \hatmbf(#1){\mathbox({\hat{\mbf({#1})}})}
\def \ninej(#1,#2,#3,#4,#5,#6,#7,#8,#9){\mathbox(\left\{\matrix
     {#1&#2&#3\cr#4&#5&#6\cr#7&#8&#9\cr}\right\})}
\def \rtov(#1,#2){\mathbox(\sqrt{{#1\over #2}})}
\def \sixj(#1,#2,#3,#4,#5,#6){\mathbox(\left\{\matrix
     {#1&#2&#3\cr#4&#5&#6\cr}\right\})}
\def \third(#1){\mathbox(\frac{#1}{3})}
\def \Trace(#1){\mathbox({\hbox{Tr} \left\{#1\right\}})}
\def \outprod(#1,#2){\mathbox({(#1\times #2)})}

\def \als{\mathbox(\alpha_s)}
\def \bra{\mathbox(\langle)}

\def \Del{\mathbox(\Delta)}

\def \heart{\mbox{$\heartsuit$}}
\def \ie{{\it i.e.}}
\def \invisible{\mbox{$\rule{0mm}{1mm}$}}
\def \ket{\mbox{$\rangle$}}
\def \lamlam{\mbox{$(\lam_i\cdot\lam_j)$}}

\def \Lam{\mbox{$\Lambda$}}
\def \lam{\mbox{$\lambda$}}
\def \rmd{{\rm d}}

\def \Sig{\mbox{$\Sigma$}}

\def \vecsig{\mathbox({\vec \sigma})}

\makeatother

\def\Del{\Delta}

\def\Voge{V_{\rm CMI}}
\def\als{\alpha_s}
\def\als{\alpha_s}

\def\ie{{\it i.e.}}
\def\lamlam{\lam_i\cdot\lam_j}
\def\lam{\lambda}

\def\rv{\vec r}
\def\sigsig{\sigv_i\cdot\sigv_j}

\newcommand{\KY}{K.~Yazaki}
\newcommand{\MO}{M.~Oka}

\renewcommand{\bra}{\langle}
\renewcommand{\half}{\mbox{$\frac{1}{2}$}}
\renewcommand{\ket}{\rangle}

\renewcommand{\rtov}[2]{\mbox{$\sqrt{\frac{#1}{#2}}$}}

\renewcommand{\third}{\mbox{$\frac{1}{3}$}}
\def\Voge{\mathbox({V_{\rm OGE}})}
\def\Vcmi{\mathbox(V_{\rm CMI})}
\def\Vconf{\mathbox(V_{\rm CONF})}

\def\forth(#1){\mathbox(\frac{#1}{4})}
\def\lamlam{\lam_i\cdot\lam_j}

\def\sigsig{\vecsig_i\cdot\vecsig_j}

\def\MO{M.~Oka} \def\KY{K.~Yazaki}

\begin{document}

\final\def\heart{}

\date{TIT/HEP-243/NP\\
      hep-ph/9312275\\
      November, 1993}

\title{Short-range YN interactions in the Quark Cluster Model%
\footnote{talk presented by M.~Oka at {\sl the JSPS-NSF Joint Seminar on
``{\it Hyperon Nucleon Interactions}''}, Maui, HI, October, 1993}}

\author{
Ken-ichiro Ogawa\thanks{e-mail: ogawa@phys.titech.ac.jp},
Sachiko Takeuchi$^{(a)}$\thanks{e-mail: sachiko@phys.titech.ac.jp}
and
Makoto Oka\thanks{e-mail: oka@phys.titech.ac.jp}\\
{\sl Department of Physics, Tokyo Institute of Technology}\\
{\sl Meguro, Tokyo 152, Japan}\\
and\\
$^{(a)}${\sl Department of Public Health and Environmental Science}\\
{\sl Tokyo Medical and Dental University}\\
{\sl Yushima, Bunkyo, Tokyo 113, Japan}}

\maketitle

\abstract {A phenomenological model for the hyperon-nucleon interactions is
constructed by using the quark cluster model approach to the
short-distance baryon-baryon interactions.  The model contains the SU(3)
symmetric meson exchange interaction at large distances and the
quark-exchange short-distance interaction.  The main feature of the model
is that strong channel dependences of the short range repulsions due to
the quark model symmetry.  It is pointed out that two channels, ($I$,
$S$)= (1/2, 0) and (3/2, 1), of the S-wave sigma-nucleon
interactions have extremely strong repulsions at short-distances.}

\thispagestyle{empty}

\newpage

\section{Introduction}
Various phenomenological models of the hyperon (Y) --  nucleon (N)
interactions are
available for studying the structure of hypernuclei{\cite{Nij,Julich}}.
They are mostly based on the meson exchange potentials with the SU(3) flavor
symmetry.
At short distances the meson-exchange potentials are supplemented by
strong repulsions, which are similar to the well-known NN repulsive core.
The models contain many arbitrary parameters concerning the
short-range repulsion, which are determined
by the very-limited experimental data of the YN scatterings.

On the other hand, the quark model description of the short-range
repulsion between two nucleons is quite successful{\cite{OY80,Nagoya}}.
It has been demonstrated in the quark
cluster model (QCM) calculations that the color-magnetic gluon
exchange and the
quark antisymmetrization in the valence quark dynamics provides a
non-local soft
repulsive core, which can reproduce the N-N scattering S matrices for
energies up to 300-400 MeV.
When QCM is applied to other two-baryon systems, the same
mechanism yields strong short-range
repulsions in most of the two ground-state baryon systems, including \Lam-N
and \Sig-N{\cite{Theo1}}.
We find that the quark
exchange effects (due to the antisymmetrization) show distinctive
spin-isospin dependences especially for the \Sig-N interactions.
Such strong channel dependences have not been considered or taken into
account in the conventional YN potential models.
Thus it is quite interesting to see whether the YN interaction model that
incorporates the quark-exchange mechanism can achieve similar successes
in explaining the YN two-body data as the conventional models.

The aim of this report is to show how one can construct a phenomenological
potential model for the hyperon-nucleon interaction incorporating both
the meson exchanges and the quark-gluon effects.   Such a model enables
us to analyze experimental YN scattering data and to determine whether
the quark-exchange mechanism is indeed at work for the hyperon-nucleon
systems.  This is important further for the study of double strange
systems, such as \Lam-\Lam, N-$\Xi$ and the H dibaryon, because the
interactions of $S=-2$ two-baryon systems are not yet directly accessible in
experiment and thus require theoretical predictions.

\section{Mechanisms of short-range repulsion in the quark model}

The short-distance repulsions between baryons seem universal for most
two-baryon interactions.
It is, for instance,  known from the study of hypernuclei that the
hyperon-nucleon interactions contain a short-range repulsion
similar to the nuclear force.
This universality can be accounted for by the simple quark model, which
provides us with two distinct mechanisms
for the short-distance repulsion{\cite{OY80,Nagoya}}.

The first (I) is due to the Pauli exclusion principle among the valence
quarks.
It can produce a strong repulsion between two baryons where quark
distributions overlap with each other.  The strength of the repulsion
can roughly be estimated from the eigenvalues of the normalization
integral kernel for the two-baryon system,
\eq
       \bra{B_1 B_2 \delta(R-S)} | {\cal A} |
             {B_1 B_2 \chi (R)} \ket  \nonumber
    &=&    \int \bra{B_1 B_2 \delta(R-S)} | {\cal A} |
            {B_1 B_2 \delta(R-S')}\ket\,\chi(S')\, dS' \nonumber\\
    &=&    \int N(S,S')\, \chi(S')\, dS'    =  e \, \chi(S)
\eq
where $\cal A$ is the quark antisymmetrization operator for all the six
quarks and $\chi(R)$ denotes the relative $B_1-B_2$ wave function.
$N(S,S')$ is called the normalization integral kernel of the resonating
group method.
The Pauli forbidden state yields $e=0$, because the antisymmetrized
state vanishes.
One obtains $e=1$, if no antisymmetrization is considered.
In general,
the eigenvalue $e$ gives a good indication of the ``forbiddeness'' of
the two-baryon system.  Namely,  if $e<1$, the channel
has a ``partially forbidden'' state and the baryonic potential has a
repulsion at short distances (or actually $R=0$).

Table 1 shows the smallest eigenvalue $e$ for various $S$--wave
YN systems.  They are evaluated for the simple harmonic
oscillator quark model.
One finds that two $N\Sig$ ($L=0$) channels, $N\Sig$ ($S=0$,
$I={1\over2}$) and $N\Sig$ ($S=1$, $I={3\over2}$), have small
eigenvalues, $e=1/9$ and $2/9$ respectively.
The corresponding eigen-function $\chi$ is the harmonic oscillator
$0s$ function, which results in $|{B_1 B_2 \chi (R)} \ket = |(0s)^6\ket$.
Thus each of these channels has an almost forbidden state, which
will cause a strong short-range repulsions{\cite{Theo1}}.

\begin{table}

\caption{
The smallest eigenvalues of the normalization kernel and the effective
core radius for various S-wave $YN$ systems.  The ``type'' indicates
the origin of the short-range repulsion, either  from the first (I) or
the second (II) mechanisms.
See the text for the effective core radius.}
\begin{center}
\begin{tabular}{|rl|c|rc|}
\hline
  $BB'$         & ($J$,$I$) &  $e$   & type & effective core radius   \\
\hline
  $N   \Lam$    & (0,${1\over2}$)& 1           &  II  & 0.40 fm     \\
  $N   \Sig$    & (0,${1\over2}$)& {$1\over9$} &  I   & 0.68 fm     \\
  $N   \Lam$    & (1,${1\over2}$)& 1           &  II  & 0.34 fm     \\
  $N   \Sig$    & (1,${1\over2}$)& 1           &  II  & 0.30 fm     \\
\hline
  $N   \Sig$    & (0,${3\over2}$)& 1           &  II  & 0.48 fm     \\
  $N   \Sig$    & (1,${3\over2}$)& {$2\over9$} &  I   & 0.67 fm     \\
\hline
\end{tabular}
\end{center}
\end{table}

The second mechanism (II) for the short range repulsion is driven by the
hyperfine interaction between quarks:
\begin{equation}
   \Vcmi = - {\als\over 4} \sum_{i<j}
              {2\pi\over 3m_im_j} \, (\lamlam)  \, (\sigsig)\,
                  \delta(\rv_{ij}) \hfill
\label{eq:CMI}
\end{equation}
which is considered to come from the color magnetic part of a gluon exchange
between quarks.
The importance of this interaction in the baryon spectrum is
manifested, for instance, in $N-\Del$, and
$\Lam-\Sig$ mass differences, and the negative
neutron mean charge square radius.

The importance of the hyperfine interaction in the
short-range $NN$ interaction has been pointed out in the quark
cluster model calculation\cite{OY80,Theo1}.
One finds that the spin-spin interaction
(\ref{eq:CMI}) produces a short-range repulsion not only for $NN$ but also
for other baryon-baryon interactions, such as $N\Lam$ and $N\Sig$.
Such calculations also indicate that the Pauli exclusion principle
(mechanism I) gives in general a
stronger short-range repulsion than the hyperfine interaction (II).

\section{Quark cluster model with the Nijmegen meson exchange potential}

The quark cluster model (QCM) is most suitable in exploring the above
mechanisms of the short-range repulsion in two-baryon systems{\cite{OY80}}.
The model incorporates the full antisymmetrization among valence
quarks and the quark exchange interactions induced by the one-gluon
exchange.
Our aim is to
construct a realistic $YN$ interaction based on QCM,
which incorporates
the quark exchange interaction at short distances and the meson
exchange potential at larger distances{\cite{OSO}}.

First, we consider a valence quark model with a hamiltonian,
\eq    H=K+\Vconf +\Voge
\eq
where $K$ is the nonrelativistic quark kinetic energy term, $\Vconf$
stands for a quark confinement potential and $\Voge$ is the
Fermi-Breit potential for the one gluon exchange.
We employ the resonating group method (RGM) wave function for the
six-quark system, given by
\eq
    \Phi_{BB'} (1\sim 6) = {\cal A} [ \phi_B(1\sim 3)\,
\phi_{B'}(4\sim6) \,\chi(R)\,]
\eq
and solve the RGM integral equation, with the kernels $H$
(Hamiltonian) and $N$ (Normalization):
\eq
  \int \left[ H(R,R') - E\, N(R,R') \right] \, \chi(R')\, dR' = 0
\label{eq:QCM}
\eq
Nonlocality of the RGM equation comes from the antisymmetrization
of the quarks.

In order to describe the long-range part of the baryon-baryon
interaction, we additionally need the meson exchange potentials.
We keep the SU(3) symmetry for the meson-baryon couplings.  Indeed,
the $YN$ potential models, such as the Nijmegen models{\cite{Nij}} and
J\"ulich models{\cite{Julich}}, are based on the SU(3) symmetry.  In
this study, we employ the
meson-exchange part of the Nijmegen potential model D and instead of
using the hard cores in the original model, superpose it with the
quark exchange interaction at the short distance.
We introduce to the QCM equation
(\ref{eq:QCM})
the meson exchange potential,
which is borrowed from the Nijmegen model D in this study.
This can be done by adding
an integral kernel for the meson exchange potential, given by
\eq
       V(R,R') \equiv \int dR'' N^{1/2}(R,R'') V_{f} (R'')
      N^{1/2}(R'',R')
\eq
where $V_{f}$ is the Nijmegen meson exchange potential with the
appropriate form factor.  The form factor is chosen so as to be consistent
with the quark wave function of the baryon,
\eq
      V_{f} (R) \equiv \int \rho(x;R/2) V_N (x-y) \rho(y;-R/2)\, dx\,dy
\eq
where $V_N$ is the original Nijmegen D potential without the repulsive
core and $\rho(x;R/2)$ stands for
the quark density of the baryon centered at $R/2$.
In the QCM calculation, we employ the Gaussian for the
internal quark wave functions of the baryon for simplicity, and thus
the corresponding form factor is given also by a Gaussian.

\section{Results}

We have five parameters in the present model: the light quark mass
$m_q$, the ratio of the light and strange quark masses  $m_q/m_s$, the
strength of linear confinement potential, $a$, the strength of the one-gluon
exchange
potential, $\alpha_s$, and
the size parameter $b$ for the Gaussian wave function of quarks in the
baryon.
In order to make the calculation consistent in kinematics, we choose
$m_q$ to be one-third of the average octet baryon mass, \ie, 383.7 MeV.
The ratio of the light/strange quark masses is fixed to 0.6, which
gives the correct $\Lam-\Sig$ mass difference.
The gluon coupling constant is chosen so as to reproduce the
$N-\Delta$ mass difference, and we also choose the
confinement $a$ so that the baryon state is stable against the breathing mode
excitation, \ie, $\partial E_B/\partial b = 0$.
The remaining parameter $b$ is sensitive to the $NN$
interaction, because it determines the size of the form factor and
also the range of the quark exchange interaction.  Therefore we leave
this as a free parameter and use the $NN$ scattering data to choose
the best value for $b$.
The QCM calculation with the Nijmegen D meson exchange potential can
fit the $NN$ $^1S_0$ scattering phase shift well for $b=0.56$ fm.
Then the other parameters are determined: $a=20.8$ MeV/fm, $\alpha_s=1.85$.
Fig.~1 shows the fit of the $NN$ $^1S_0$ scattering phase shift
calculated with this choice of parameters.
{\heart}The fit is not very good for low energy.  In order to improve the fit,
we might adjust some of the meson exchange parameters, which at the
present are taken from the Nijmegen D model.

\begin{figure}
\begin{center}
  \unitlength=1cm
  \begin{picture}(6,1)
   \put(0,0){\framebox(6,1){FIGURE 1 [NN1s0.ps]}}
  \end{picture}
\end{center}
\caption{$^1S_0$ NN scattering phase shifts.}
\end{figure}

We then calculate the scattering S matrices for various  $YN$ systems in
this model and
find that the qualitative predictions given above are
confirmed in the present model.
Fig.~2 shows the $\Lambda N$ $^1S_0$ scattering phase shifts.
The result is compared with that for the original Nijmegen model D with
the hard core.  One sees that the $\Lambda N$ interaction in QCM is
more attractive than that in the original Nijmegen model.
This channel couples with the $\Sigma N$ $^1S_0$ state.  The effect of
the coupling is significant especially for QCM.

\begin{figure}
\begin{center}
  \unitlength=1cm
  \begin{picture}(6,1)
   \put(0,0){\framebox(6,1){FIGURE 2 [NLS1s0.ps]}}
  \end{picture}
\end{center}
\caption{$^1S_0$ $N\Lambda$ scattering phase shifts.  The dotted and the
dash-dotted curves are for the calculations without couplings to
$N\Sigma$.}
\end{figure}

Figs.~3 and 4 show the $\Sigma N$ scattering phase shifts for
$I={1\over 2}$, $^1S_0$ and $I={3\over 2}$, $^3S_1$ channels.  These are
the channels where the Pauli principle expects the type I strong
repulsion.
We indeed obtain strongly repulsive phase shifts, far more repulsive
than the original Nijmegen model,
while QCM yields milder repulsions in the other $\Lambda N$ and
$\Sigma N$ channels.
Therefore these two Pauli-forbidden $\Sigma-N$ interactions are
exceptional.
This indicates that the $\Sigma-N$ interactions have  strong spin-isospin
dependences.

\begin{figure}
\begin{center}
  \unitlength=1cm
  \begin{picture}(6,1)
   \put(0,0){\framebox(6,1){FIGURE 3 [NSL1s0.ps]}}
  \end{picture}
\end{center}
\caption{$I={1\over 2}$ $^1S_0$ $N\Sigma$ scattering phase shifts with
couplings to $N\Lambda$.}
\end{figure}

\begin{figure}
\begin{center}
  \unitlength=1cm
  \begin{picture}(6,1)
   \put(0,0){\framebox(6,1){FIGURE 4 [NS3s1.ps]}}
  \end{picture}
\end{center}
\caption{$I={3\over 2}$ $^3S_1$ $N\Sigma$ scattering phase shifts.}
\end{figure}

In Table 1,  the properties of the short-distance
$YN$ interactions are summarized in terms of the effective size of the
repulsive core, defined by
$d\delta/dk$ at $E_{lab}=350$ MeV.
One sees that the Pauli exclusion principle gives a stronger repulsion
for the $N\Sig$ ($S=0$, $I={1\over2}$) and
$N\Sig$ ($S=1$, $I={3\over2}$) channels,
while the other channels show a mild repulsion which is generally
softer than the original Nijmegen model D.
The repulsion in the $N\Sig$ ($I={3\over 2}$, $^3S_1$) channel is as
strong as that in the Nijmegen model F, which is known to provide not enough
binding for $\Sig$ to make a bound $\Sig$ hypernuclei.
Details of the model and the results will be published
elsewhere{\cite{OSO}}.

\section{Conclusion and Discussion}

We present a quark model analysis of hyperon-nucleon interactions.
The realistic  $YN$ interactions, which are nonlocal due to the
quark antisymmetrization effects, are proposed using the quark cluster
model approach with the Nijmegen model D meson exchange potential.
The main difference between the original Nijmegen model and our
interaction arises in the spin-isospin dependence of the $YN$ short
range interactions.  Especially, $\Sig N$ with $S=0$, $I=1/2$ and
$S=1$ and $I=3/2$ have strong repulsion at the short distance in the
quark model and may make the bound $\Sig$ hypernuclei implausible.

{\heart}The model is not complete yet.  The final goal is a
no-parameter model for the YN (and YY) interactions based on the SU(3)
symmetry for the meson-baryon coupling constants.
The quark model parameters and the meson-baryon coupling for each SU(3)
multiplet can be determined in the non-strange sector, that is the NN
sactterings.
The present model borrows the meson-baryon couplings from the Nijmegen
potential, whose short-range behaviors are different from the quark
cluster model. Thus the fit to the NN data is not as complete as the
original potential.
One has to adjust the coupling constants so as to reproduce the NN
data.  Work along this line in under way.

\end{document}